\documentclass[showpacs,aps,prd,reprint,superscriptaddress,nofootinbib]{revtex4-1}
\usepackage[colorlinks=true, pdfstartview=FitV, linkcolor=magenta,citecolor=blue, urlcolor=blue,
bookmarks=true, bookmarksnumbered=true, breaklinks]{hyperref}
\usepackage[dvipdfmx]{graphicx}

\usepackage{amsmath,amssymb,bm,color,longtable,mathrsfs,slashed}

\begin{document}
\begin{flushright}
KEK-TH-1993
\end{flushright}

\title{Interplay between chiral symmetry breaking and the QCD Kondo effect}

\author{Kei~Suzuki}
\email[]{{\tt kei.suzuki@kek.jp}}
\affiliation{KEK Theory Center, Institute of Particle and Nuclear
Studies, High Energy Accelerator Research Organization, 1-1 Oho, Tsukuba, 
Ibaraki, 305-0801, Japan}
\author{Shigehiro~Yasui}
\email[]{{\tt yasuis@th.phys.titech.ac.jp}}
\affiliation{Department of Physics, Tokyo Institute of Technology, Tokyo 152-8551, Japan}
\author{Kazunori~Itakura}
\email[]{{\tt kazunori.itakura@kek.jp}}
\affiliation{KEK Theory Center, Institute of Particle and Nuclear
Studies, High Energy Accelerator Research Organization, 1-1 Oho, Tsukuba, 
Ibaraki, 305-0801, Japan}
\affiliation{Graduate University for Advanced Studies (SOKENDAI),
1-1 Oho, Tsukuba, Ibaraki, 305-0801, Japan}

\begin{abstract}
We investigate the interplay between the (light-light quark) chiral condensate and heavy-light quark condensate induced by the QCD Kondo effect, which is described by an effective Lagrangian with four-point interactions and the mean-field approximation. 
We find that the appearance of Kondo condensates decreases the critical chemical potential of the chiral condensate.
In the region near the critical chemical potential, a coexistence phase with two kinds of condensates can appear.
The behavior of such an interplay at finite temperature and with nonzero current light-quark mass is also discussed.
\end{abstract}

\pacs{12.39.Hg,21.65.Qr,12.38.Mh,72.15.Qm}
\keywords{Quark matter, Nuclear matter, Kondo effect, Heavy quark effective theory}

\maketitle

\section{Introduction}
The Kondo effect is one of the most well-known subjects in condensed matter physics: it manifests itself as transport properties of conduction electrons in a metal, including localized heavy impurities \cite{Kondo:1964,Hewson,Yosida,Yamada}.
In hadron and nuclear physics---which should be described by QCD---similar situations appear, such as the isospin Kondo effect \cite{Yasui:2013xr,Yasui:2016ngy} and QCD Kondo effect \cite{Yasui:2013xr,Hattori:2015hka,Ozaki:2015sya,Yasui:2016svc,Yasui:2016yet,Kanazawa:2016ihl,Kimura:2016zyv,Yasui:2017izi}, which are mediated by isospin- and color-exchange interactions, respectively.\footnote{Of course, even the spin-exchange interactions between a light hadron and a heavy hadron can induce the Kondo effect \cite{Yasui:2016hlz}.}
These new types of Kondo effects can be related to the transport and thermodynamic properties of dense nuclear/quark matter with heavy impurities.
Thus, they are relevant for realistic observables in compact stars and heavy-ion collisions at FAIR, NICA, and J-PARC, where nuclear/quark matter exists and heavy hadrons/quarks can be created through high-energy processes such as cosmic neutrino flux and nucleon-nucleon collisions.

The Kondo effect in a simple situation can be described by perturbation theory.
It is derived from loop effects in relatively high-energy scattering between a light fermion and a heavy impurity; then, the infrared divergence of the scattering amplitude indicates the drastic change in the transport properties (e.g., electric conductivity). 
On the other hand, to describe the ``nonperturbative" region of the Kondo effects is a somewhat challenging problem, where one has to utilize some theoretical techniques such as mean-field approximations.

In our previous works \cite{Yasui:2016svc,Yasui:2017izi}, we investigated the QCD Kondo effect by focusing on the roles of the color-exchange interaction between a light quark and a heavy quark impurity.
As a mean-field approximation for the QCD Kondo effect, we suggested the heavy-light quark condensate ({\it Kondo condensate}) as a color-singlet ground state of a quark matter which contains heavy-quark impurities.
Although the Kondo condensate can appear in a wide range of light-quark chemical potentials \cite{Yasui:2016svc,Yasui:2017izi}, it would be affected by other QCD effects which were not considered in the previous works.
In Ref.~\cite{Kanazawa:2016ihl}, the competition with the color superconducting phase \cite{Alford:2007xm}---which can be realized in high-density quark matter---was discussed.
On the other hand, at lower density near the confining phase, its behavior is still unknown.
In fact, in such a region a condensate composed of a light quark and its light antiquark (that is, a chiral condensate) is realized, and it is the most essential property of the low-energy region of QCD \cite{Nambu:1961tp,Nambu:1961fr}.

In this paper, to focus on the QCD Kondo effect in {\it low-density} quark matter, we investigate the interplay between the usual chiral condensate and the Kondo condensate.
We will show that the Kondo condensate with a strong enough heavy-light interaction can drastically affect the chiral condensate, and our main findings are the {\it exclusion effect} of the chiral condensate and the {\it appearance of a coexistence phase} between the condensates.
Such an investigation could be useful for the search for the chiral phase transition through medium modification of heavy hadrons---such as $D$, $D_s$, and $B$ mesons in nuclear matter \cite{Hilger:2008jg,Blaschke:2011yv,Hilger:2011cq,Sasaki:2014asa,Buchheim:2014rpa,Suenaga:2014sga,Suzuki:2015est,Park:2016xrw,Harada:2016uca,Suenaga:2017deu} (see Ref.~\cite{Hosaka:2016ypm} for a recent review)---and for the evaluation of the ``nonperturbative" aspect of the QCD Kondo effect near the critical density.

This paper is organized as follows.
In Sec.~\ref{Sec_2}, we formalize the mean-field approach to describe both the Kondo and chiral condensates.
In Sec.~\ref{Sec_3}, the numerical results for the phase diagram of quark matter including the two kinds of condensates are shown.
Section~\ref{Sec_4} is devoted to our conclusion and perspective.

\section{Formalism} \label{Sec_2}
\subsection{Mean-field Lagrangian}
The Lagrangian includes light-quark fields $\psi \equiv(\psi_1^t, \psi_2^t, \cdots, \psi_{N_f}^t)$ and a heavy-quark field $\Psi_v \equiv \frac{1}{2} (1 + v^\mu \gamma_\mu) e^{im_Q v \cdot x} \Psi$, where $m_Q$ and $v^\mu =(1,\vec{0})$ are the heavy-quark mass and heavy-quark four-velocity at rest, respectively \cite{Eichten:1989zv,Georgi:1990um} (see Refs.~\cite{Neubert:1993mb,Manohar:2000dt} for reviews).
The Lagrangian with the color-current interaction inspired by QCD is given by
\begin{eqnarray}
{\cal L} 
&=&\bar{\psi} i \partial\hspace{-0.55em}/ \psi + \mu \, \bar{\psi} \gamma^{0} \psi - G_{qq} \sum_{a} \left( \bar{\psi} \gamma^{\mu} T^{a} \psi \right) \left( \bar{\psi} \gamma_{\mu} T^{a} \psi \right) \nonumber \\
&& + \bar{\Psi}_v i v^\mu \partial_\mu \Psi_v - \lambda (\bar{\Psi}_v \Psi_v - n_Q ) \nonumber \\ 
&& - G_{Qq} \sum_{a} \left( \bar{\psi} \gamma^{\mu} T^{a} \psi \right) \left( \bar{\Psi}_v \gamma_{\mu} T^{a} \Psi_v \right),
\label{eq:L}
\end{eqnarray}
where $T^a = \lambda^a/2$ are the generators of color SU($N_c$) and the color indices ($a=1, \dots, N_c^2-1$) are summed over.
After the Fierz transformation, we obtain\footnote{To derive the heavy-light sector of the Lagrangian, we used the relation $\gamma^0 \Psi_v = \Psi_v$ which comes from the heavy-quark limit.
By this relation, the vector-current interaction is rewritten in the following form:
\begin{equation}
 ( \bar{\psi} \gamma_\mu \Psi_v ) (\bar{\Psi}_v \gamma^\mu \psi) = ( \bar{\psi} \Psi_v ) (\bar{\Psi}_v \psi) - ( \bar{\psi} \vec{\gamma} \Psi_v ) (\bar{\Psi}_v \vec{\gamma} \psi). \nonumber
\end{equation}
}
\begin{eqnarray}
{\cal L} 
&=&\bar{\psi} i \partial\hspace{-0.55em}/ \psi + \mu \, \bar{\psi} \gamma^{0} \psi + \frac{G_{qq} (N_c^2 -1)}{2N_f N_c^2}  (\bar{\psi} \psi )^2  \nonumber \\
&& + \bar{\Psi}_v i v^\mu \partial_\mu \Psi_v - \lambda (\bar{\Psi}_v \Psi_v - n_Q ) \nonumber \\ 
&& + \frac{G_{Qq} (N_c^2 -1)}{4 N_c^2} \left[ ( \bar{\psi} \Psi_v ) (\bar{\Psi}_v \psi) + ( \bar{\psi} \vec{\gamma} \Psi_v )( \bar{\Psi}_v \vec{\gamma} \psi ) \right] \nonumber\\
&& + \cdots, \label{eq:L2}
\end{eqnarray}
where $N_f$ and $N_c$ are the number of flavors and colors of light quarks, respectively.
For the light-light sector, we apply the Nambu--Jona-Lasinio (NJL)-type four-point interaction with the coupling constant $G_{qq}>0$  \cite{Vogl:1991qt,Klevansky:1992qe,Hatsuda:1994pi,Buballa:2003qv}.
For the heavy-light sector, as suggested in our previous works \cite{Yasui:2016svc,Yasui:2017izi}, we assume the color-singlet interaction with $G_{Qq} >0$, which can be justified in the large-$N_c$ limit.
$\mu$, $\lambda$, and $n_Q$ are the light-quark chemical potential, Lagrange multiplier,\footnote{In the framework of heavy-quark effective theory (HQET) \cite{Eichten:1989zv,Georgi:1990um} with $v^\mu =(1,\vec{0})$, because all of the heavy quarks are at rest in position space, they do not have a phase space of real (on-shell) momentum.
Therefore, the heavy quarks cannot form a Fermi sphere in momentum space.
Of course, since $-\lambda$ is the coefficient of $\bar{\Psi}_v \Psi_v = \Psi^\dag_v \Psi_v$, one may formally interpret it as the chemical potential of heavy quarks.
Then, it is the ``chemical potential" for the {\it redefined} heavy-quark field $\Psi_v$ rather than the original heavy-quark field $\Psi$.
In other words, a nonzero $\lambda$ can be regarded as the energy necessary to put a virtual component (residual momentum) of a heavy quark into the system.} and heavy-quark number density, respectively.
This model leads to spontaneous chiral-symmetry breaking through the light-light quark interaction with $G_{qq}$, and the formation of the Kondo condensate through the heavy-light quark interaction with $G_{Qq}$.

We introduce the mean-field approximation for the chiral condensate, and the scalar and vector heavy-light condensates for massive light flavors as follows:
\begin{eqnarray}
\langle \bar{\psi} \psi \rangle &\equiv& - \frac{ N_f N_{c}^2}{(N_c^2-1)G_{qq}} M, \label{MFA1} \\ \nonumber \\
\langle \bar{\psi}_i \Psi_v \rangle &\equiv& \frac{ 4N_{c}^2}{(N_c^2-1)G_{Qq}} \Delta \sqrt{\frac{E_p + M}{E_p}}, \label{MFA2} \\ 
\langle \bar{\psi}_i \, \vec{\gamma} \, \Psi_v \rangle &\equiv& \frac{ 4N_{c}^2}{(N_c^2-1)G_{Qq}}  \Delta \sqrt{\frac{E_p + M}{E_p}} \frac{\vec{p} \cdot \vec{\gamma}}{E_p + M}, \label{MFA3}
\end{eqnarray}
where $E_p=\sqrt{p^2+M^2}$, and $M$ and $\Delta$ are the constituent quark mass (or chiral condensate) for light flavors and the gap of the Kondo condensate, respectively.
Here, the $i$ light flavors $\psi_i$ couple with the heavy flavor.
In this work, we assume $i=1$ for $N_f=2$.
Note that, by taking $M \to 0$, we can confirm that this form agrees with the case of massless light flavors in the previous work \cite{Yasui:2017izi}.

From the original Lagrangian~(\ref{eq:L2}) and the mean-field approximations~(\ref{MFA1})--(\ref{MFA3}), we can write down the mean-field Lagrangian.
Here, it is convenient to define the following notation:
\begin{eqnarray}
\phi \equiv 
\left( \!\!\!
\begin{array}{c}
 \psi_{1} \\
 \psi_{2} \\
 \vdots \\
 \psi_{N_f} \\
 \Psi_{v}^\prime  
\end{array}
\!\!\! \right),
\hspace{2em}
 \bar{\phi} \equiv (\bar{\psi}_{1}, \bar{\psi}_{2}, \dots, \bar{\psi}_{N_f}, \bar{\Psi}_{v}^\prime),
 \label{eq:phi_def}
\end{eqnarray}
where the heavy-quark field $\Psi_v$ was projected to the positive-energy (two-spinor) component $\Psi_v^\prime$ defined by
\begin{eqnarray}
\Psi_{v} \equiv 
\left( \!\!\!
\begin{array}{c}
 \Psi_{v}^\prime \\
 0
\end{array}
\!\!\! \right),
\hspace{2em}
 \bar{\Psi}_v \equiv (\bar{\Psi}_v^{\prime}, 0).
\end{eqnarray}
Using the notation (\ref{eq:phi_def}), the mean-field Lagrangian is rewritten in the following compact form:
\begin{eqnarray}
{\cal L}_{\mathrm{MF}} &=& \bar{\phi} \,  G(p_{0},\vec{p}\,)^{-1} \phi -\frac{8N_{c}^{2}}{(N_{c}^{2}-1)G_{Qq}} |\Delta|^{2} \nonumber\\
&& - \frac{N_f N_c^2}{2(N_c^2-1)G_{qq}} M^2 + \lambda n_Q, \label{eq:MF}
\end{eqnarray}
where $G(p_{0},\vec{p}\,)^{-1}$ is the inverse propagator for the field $\phi$.

\begin{widetext}{
\subsection{$N_f=1$ case}
For $N_f=1$, the inverse propagator is given by
\begin{eqnarray}
 G(p_{0},\vec{p}\,)^{-1}
\equiv
 \left(
\begin{array}{ccc}
p_0 + \mu - M             & -\vec{p}\cdot \vec{\sigma} & \Delta^\ast \sqrt{\frac{E_p + M}{E_p}} \\
\vec{p}\cdot \vec{\sigma} & -(p_0 + \mu) - M           & -\Delta^\ast \sqrt{\frac{E_p + M}{E_p}} \frac{\vec{p}\cdot \vec{\sigma}}{E_p + M} \\
\Delta \sqrt{\frac{E_p + M}{E_p}}  & \Delta  \sqrt{\frac{E_p + M}{E_p}} \frac{\vec{p}\cdot \vec{\sigma}}{E_p + M} & p_0 - \lambda \\
\end{array}
\right),
\label{eq:Ginverse_RL_Nf}
\end{eqnarray}
where the $3 \times 3$ matrix is the two-component form of Dirac gamma matrices and the one component of the heavy-quark field.
By taking $\det G(p_{0},\vec{p}\,)^{-1} =0$, we obtain the dispersion relation for $N_f=1$:
\begin{eqnarray}
E_\pm(p) &\equiv&  \frac{1}{2} \left( \sqrt{p^2 + M^2} + \lambda -\mu \pm \sqrt{\left(\sqrt{p^2 + M^2}-\lambda-\mu \right)^2 + 8 |\Delta|^2 }\right), \label{dis_Epm_Nf=1} \\
\tilde{E}(p) &\equiv& \tilde{E}_{1} = - \sqrt{p^2 + M^2} - \mu. \label{dis_Etil_Nf=1}
\end{eqnarray}

\subsection{$N_f \geq 2$ case}
For $N_f\geq 2$, we assume that only the first light flavor $\psi_1$ couples to the heavy flavor, as we mentioned before [see Eqs.~(\ref{MFA2}) and (\ref{MFA3})].\footnote{
As another choice for $N_f\geq 2$, we may couple both $\psi_1$ and $\psi_2$ to the heavy flavor $\Psi_v$.
Then we obtain the following dispersion relation for heavy-light mixing modes:
\begin{eqnarray}
E_\pm(p) &\equiv&  \frac{1}{2} \Biggr( \sqrt{p^2 + M^2} + \lambda -\mu \Biggr. \nonumber \\
&& \left.\pm \sqrt{ \left( \sqrt{p^2 + M^2}-\lambda-\mu \right)^2 + 8N_f |\Delta|^2 }\right).
\end{eqnarray}
This setup for massless light flavors was used in our previous work \cite{Yasui:2016svc}.
We emphasize that such different definitions of the Kondo condensate lead to the same minimum of the free energy of the system.
In other words, all of the choices for $N_f=2$, $\langle \bar{\psi}_1 \Psi_v \rangle$, $\langle \bar{\psi}_2 \Psi_v \rangle$, and $\langle \bar{\psi}_1 \Psi_v \rangle$ + $\langle \bar{\psi}_2 \Psi_v \rangle$, are degenerate as the ground state. 
In fact, in real QCD, since the up and down quarks are slightly different from each other, either of the light flavors could be favored by slight differences, such as the current quark mass and electric charge.
In such a case, the Kondo condensate in which only one light flavor couples to the heavy quark (as formalized in the main text) will be useful.
}
Then the inverse propagator is given by
\begin{eqnarray}
 G(p_{0},\vec{p}\,)^{-1}
\equiv
\left(
\begin{array}{cccccc}
p_0 + \mu - M             & -\vec{p}\cdot \vec{\sigma} & 0 & 0 & \hdots & \Delta^\ast \sqrt{\frac{E_p + M}{E_p}} \\
\vec{p}\cdot \vec{\sigma} & -(p_0 + \mu) - M           & 0 & 0 & \hdots & -\Delta^\ast \sqrt{\frac{E_p + M}{E_p}} \frac{\vec{p}\cdot \vec{\sigma}}{E_p + M}\\
0  & 0 & p_0 + \mu - M      & -\vec{p}\cdot \vec{\sigma} & \hdots & 0 \\
0  & 0 & \vec{p}\cdot \vec{\sigma} & -(p_0 + \mu) - M    & \hdots & 0 \\
\vdots & \vdots  &  \vdots  & \vdots &  \ddots & \vdots \\ 
\Delta \sqrt{\frac{E_p + M}{E_p}}  & \Delta  \sqrt{\frac{E_p + M}{E_p}} \frac{\vec{p}\cdot \vec{\sigma}}{E_p + M} & 0 & 0 & \hdots & p_0 - \lambda \label{detNf=2} \\
\end{array}
\right).
\label{eq:Ginverse_RL_Nf2}
\end{eqnarray}
By taking $\det G(p_{0},\vec{p}\,)^{-1} =0$, we obtain the energy-momentum dispersion relations for $N_f$ flavors as follows:
\begin{eqnarray}
E_\pm(p) &\equiv&  \frac{1}{2} \left( \sqrt{p^2 + M^2} + \lambda -\mu \pm \sqrt{\left(\sqrt{p^2 + M^2}-\lambda-\mu \right)^2 + 8 |\Delta|^2 }\right), \label{dis_Epm_Nf=2}\\
E(p) &\equiv& E_i = \sqrt{p^2 + M^2} - \mu \hspace{17pt} [i=2, \cdots, N_f], \label{dis_E_Nf=2} \\
\tilde{E}(p) &\equiv& \tilde{E}_i = - \sqrt{p^2 + M^2} - \mu \hspace{10pt} [i=1, \cdots, N_f]. \label{dis_Etil_Nf=2}
\end{eqnarray}

\subsection{Thermodynamic potential}
The thermodynamic potential at finite temperature $T$ is given by
\begin{equation}
\Omega(T,\mu,\lambda;\Delta,M) = 2N_c \int_0^\Lambda f(T,\mu,\lambda;p) \frac{p^2 dp}{2\pi^2} + \frac{8N_c^2}{(N_c^2-1)G_{Qq}} |\Delta|^2 + \frac{N_f N_c^2}{2(N_c^2-1)G_{qq}} M^2 - \lambda n_Q,   \label{omega} 
\end{equation}
where the factor $2N_c$ in front of the integral comes from the chirality and color for the light quarks, and
\begin{equation}
f(T,\mu,\lambda;p) = -\frac{1}{\beta} \ln \left[ (1+e^{-\beta E_+(p)}) (1+e^{-\beta E_-(p)}) (1+e^{-\beta E(p)})^{N_f-1} (1+e^{-\beta \tilde{E}(p)})^{N_f} \right].
\end{equation}

}\end{widetext}
From the minimization of Eq.~(\ref{omega}) at fixed $T$, $\mu$, and $\lambda$, we can obtain the values of $\Delta$ and $M$.

Finally, we set the parameters as $G_{qq} = 2(9/2) / \Lambda^{2}$, $\Lambda = 0.65 \ \mathrm{GeV}$, and $N_{c}=3$, as usual in NJL model studies of light flavors with $N_f=2$ \cite{Klevansky:1992qe}.
These parameters were determined so as to reproduce the experimental values such as the pion mass and pion decay constant in vacuum.
On the other hand, we do not know parameters for the heavy-light sector because the heavy-light coupling constant $G_{Qq}$ at finite density can be drastically changed by the QCD Kondo effect.\footnote{As an alternative approach, the heavy-light coupling constant {\it in vacuum} can be determined from the experimental values for $D$ mesons in vacuum as in Refs.~\cite{Ebert:1994tv,Mota:2006ex,Blaschke:2011yv,Guo:2012tm}.
However, in this work we do not use such parameters to focus on the QCD Kondo effect at finite density.
In fact, since we do not know reliable values of the coupling constants $G_{Qq}$ and $G_{qq}$ at finite density, we cannot compare these coupling constants.
Naively, the typical interaction range of light-light quark scattering is larger than that of heavy-light scattering.
Therefore, the former is more strongly affected by finite-density effects including Pauli blocking between light quarks, so that the coupling constant $G_{qq}$ in the effective model can be smaller than $G_{Qq}$ .
Thus, the situation of $G_{Qq} > G_{qq}$ at finite density might be realized.}
In this work, we treat $G_{Qq}$ as a free parameter.

\section{Numerical results} \label{Sec_3}
\subsection{Results at $T=\lambda=0$}
The numerical results for $\mu$ dependences of chiral and Kondo condensates at $T=0$ and $\lambda=0$ are shown in Fig.~\ref{dense_dep}.
We show the results for both the $N_f=1$ and $N_f=2$ cases, where the value of the heavy-light coupling constant is controlled as $G_{Qq}=2G_{qq}$, $G_{qq}$ or $0.5G_{qq}$.
From these figures, our findings are as follows:

\begin{figure}[b!]
    \begin{minipage}[t]{1.0\columnwidth}
        \begin{center}
            \includegraphics[clip, width=1.0\columnwidth]{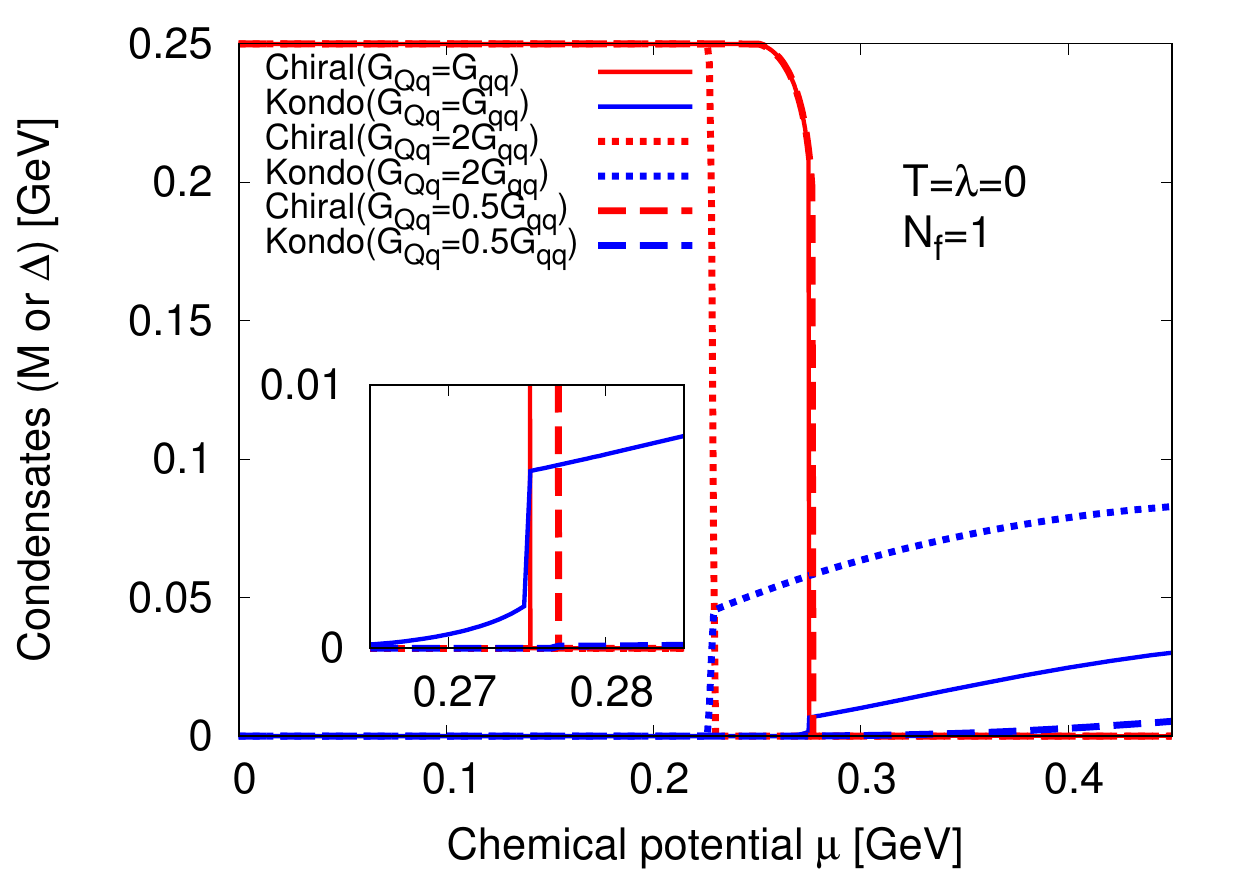}
        \end{center}
    \end{minipage}
    \begin{minipage}[t]{1.0\columnwidth}
        \begin{center}
            \includegraphics[clip, width=1.0\columnwidth]{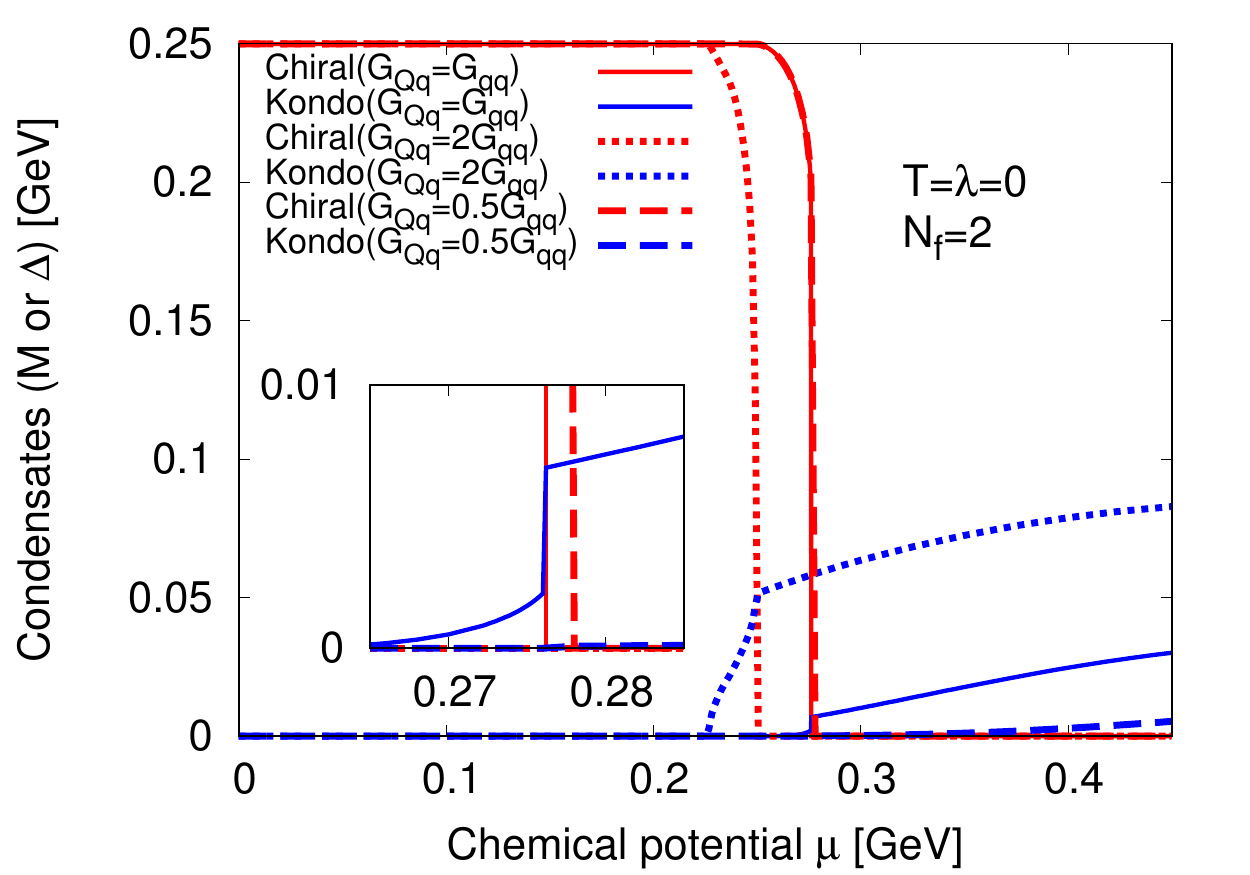}
        \end{center}
    \end{minipage}
    \caption{$\mu$ dependences of chiral and Kondo condensates at $T=\lambda=0$.
The light-quark flavor is set as $N_f=1$ (upper) or $N_f=2$ (lower).
The inset figures show the behaviors around the critical chemical potentials for the cases with $G_{Qq}=G_{qq}$ and $0.5G_{qq}$.}
    \label{dense_dep}
\end{figure}

\begin{figure}[b!]
    \begin{minipage}[t]{1.0\columnwidth}
        \begin{center}
            \includegraphics[clip, width=1.0\columnwidth]{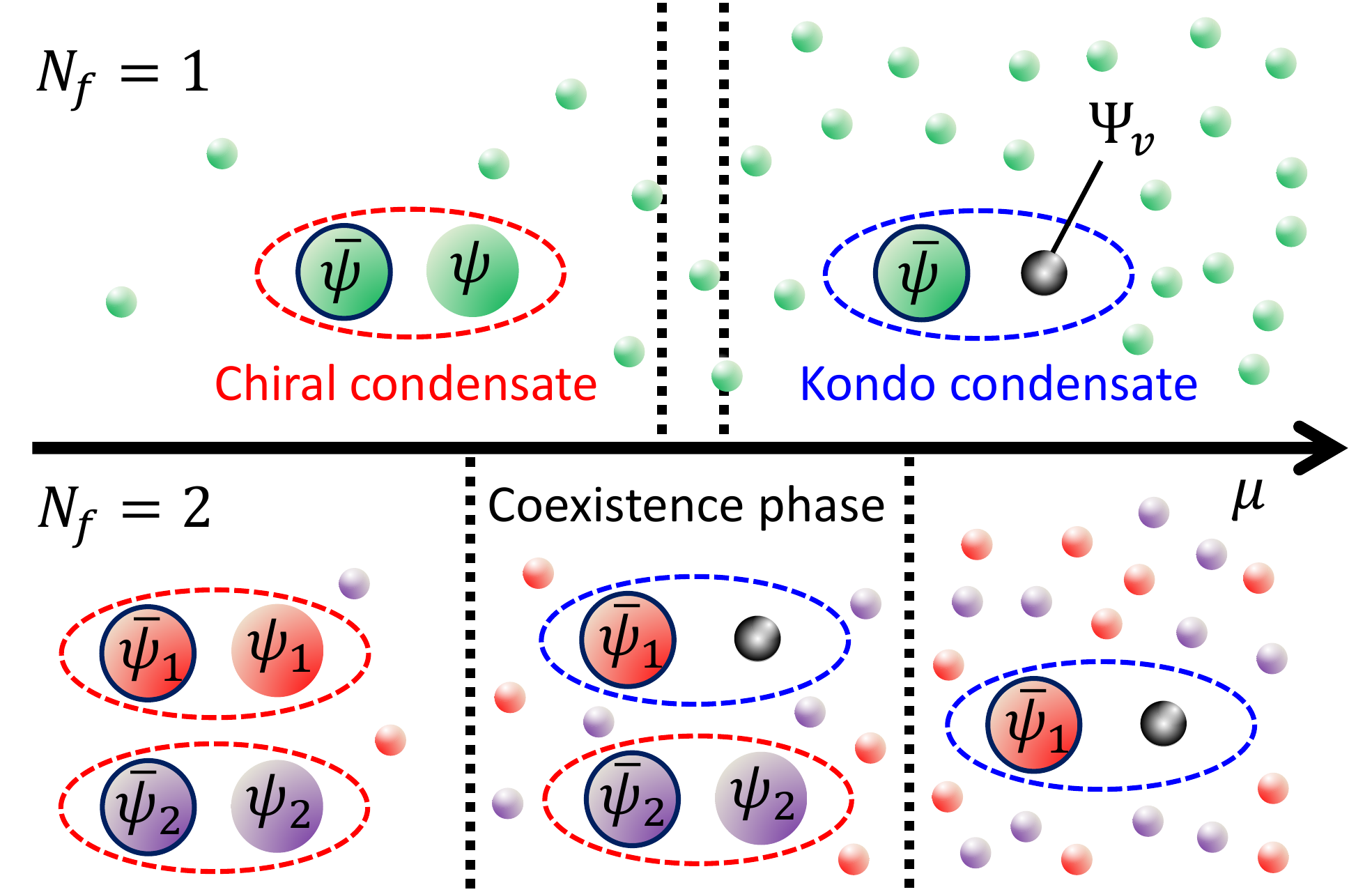}
        \end{center}
    \end{minipage}
    \caption{Schematic picture of the difference between $N_f=1$ and $N_f=2$.
Black and other colored circles denote heavy quark $\Psi_v$ and light flavors $\psi_i$, respectively.
}
    \label{flavor}
\end{figure}

\begin{enumerate}
\setlength{\leftskip}{1.2em}
\item The critical chemical potential of the chiral condensate $\mu_c^\chi$, which characterizes the chiral phase transition at finite density, is pushed down by the Kondo condensate.
For instance, when $G_{Qq}=0$ and the Kondo condensate cannot be formed, we find $\mu_c^\chi = 0.2765 \ \mathrm{GeV}$, which is nothing but the result from the usual NJL model.
With increasing $G_{Qq}$, the magnitude of the Kondo condensate becomes larger, and $\mu_c^\chi$ becomes smaller.
At $G_{Qq}=2G_{qq}$, we see that $\mu_c^\chi$ decreases by a few tens of MeV as shown in Fig.~\ref{dense_dep}.
Therefore, we conclude that the chiral condensate near the critical chemical potential is inhibited by the presence of a strong Kondo effect.

\item The coexistence phase of chiral and Kondo condensates appears near $\mu_c^\chi$.
Here, such a phase corresponds to the nonzero values of $\Delta$ and $M$ in the minimum of the thermodynamic potential $\Omega(\mu;\Delta,M)$ at a value of $\mu$.
For $G_{Qq}=G_{qq}$, the impact of such a phase is not so large, and the phase transitions of the chiral and Kondo condensates occur at almost the same $\mu$.

\item We discuss the difference between $N_f=1$ and $N_f=2$.
In particular, for $N_f=2$, the exclusion of the chiral condensate is weaker than in the $N_f=1$ case.
In other words, the critical chemical potential shows $\mu_c^\chi(N_f=1) < \mu_c^\chi(N_f=2)$.
Moreover, for $N_f=2$, the coexistence phase exists in a wider range of $\mu$ than in the $N_f=1$ case.
An intuitive picture is given as follows (see also Fig.~\ref{flavor}).
For $N_f=2$, we assumed that only the first light flavor $\psi_1$ couples with the heavy flavor $\Psi_v$.
Therefore, its chiral condensate, $\langle \bar{\psi}_1 \psi_1 \rangle$, can be excluded by the Kondo condensate, $\langle \bar{\psi}_1 \Psi_v \rangle$, while the chiral condensate for another light flavor, $\langle \bar{\psi}_2 \psi_2 \rangle$, is not affected.
As a result, the Kondo condensate for the first light flavor can coexist with the chiral condensate for the second one, which can induce the wider range of the coexistence phase.

\item If we neglect $M$ at $T=\lambda=0$, the Kondo condensate appears through the second-order phase transition with increasing $\mu$ \cite{Yasui:2016svc,Kanazawa:2016ihl}.
For finite $M$, the order of the transition to the coexistence phase is second order, and that to the Kondo phase becomes first order.
\end{enumerate}

\begin{figure}[b!]
    \begin{minipage}[t]{1.0\columnwidth}
        \begin{center}
            \includegraphics[clip, width=1.0\columnwidth]{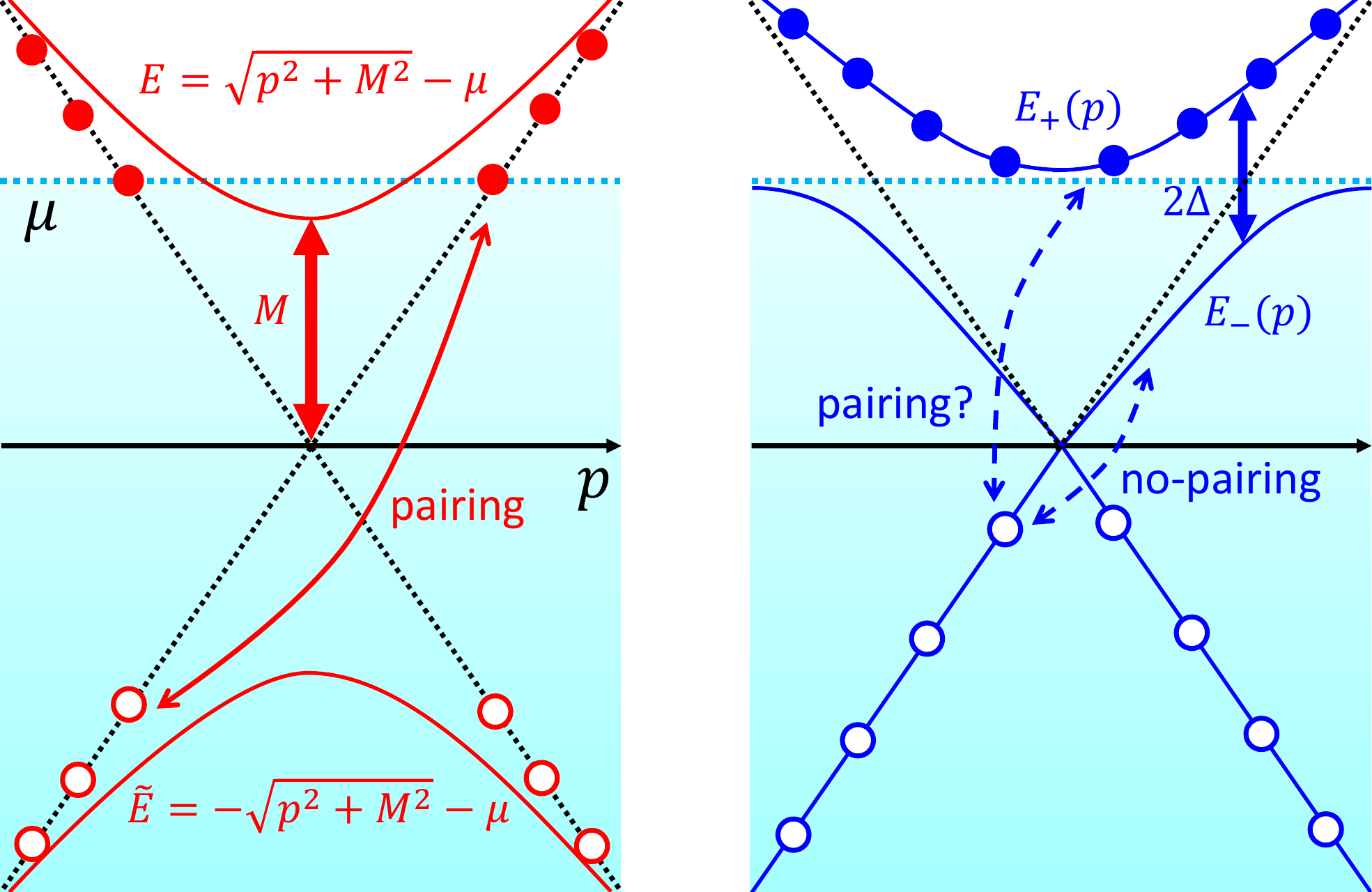}
        \end{center}
    \end{minipage}
    \caption{
Schematic pictures of dispersion relations.
Left: Dispersions of a massive quark and antiquark, $E(p)=\sqrt{p^2+M^2}-\mu$ and $\tilde{E}(p)=-\sqrt{p^2+M^2}-\mu$, with a gap $M$ at large $\mu$ (red solid lines). The filled and open circles on the dotted lines denote massless quarks and antiquarks participating in the formation of the chiral condensate, respectively.
The thin arrow denotes an example of the pairing between a quark and antiquark, which forms the chiral condensate.
Right: Dispersions of heavy-light mixed modes, $E_-(p)$ and $E_+(p)$, induced by the Kondo condensate with a gap $\Delta$ (blue solid lines).
The filled and open circles on the solid lines denote an upper mode and antiquark, respectively, which can participate in the formation of the chiral condensate.
The thin dashed arrows denote examples of the possible (impossible) pairing between an upper (lower) mode and antiquark, where the chiral condensate is disfavored (see the text).
}
    \label{dirac_cone}
\end{figure}

At the end of this subsection, we provide an intuitive picture of the exclusion effect on the chiral condensate by the Kondo condensate.
In the absence of the Kondo condensate, the chiral condensate at large $\mu < \mu_c^\chi$ is formed by the pairing between a quark near the Fermi surface and its antiquark, as shown in the left panel of Fig.~\ref{dirac_cone}.
As a result, the massless quark and its antiquark obtain an effective mass $M$, and their dispersion relations become $E(p)$ and $\tilde{E}(p)$, respectively.
Furthermore, at $\mu > \mu_c^\chi$, a nonzero chiral condensate (or $M$) becomes disfavored by the term proportional to $M^2$ in the thermodynamic potential (\ref{omega}).

Next, when the Kondo condensate appears and the gap $\Delta$ becomes finite, the dispersion of the massless light quark is separated into the lower and upper modes, $E_-(p)$ and $E_+(p)$, as quasi- (or mixed) particles by the coupling with a heavy quark, as shown in the right panel of Fig.~\ref{dirac_cone}.
Then the dispersion of the lower mode is inside the Fermi sphere, where the particles cannot form the chiral condensate due to Pauli blocking.
At the same time, the dispersion of the upper mode is lifted above the Fermi surface by the mixing (or level repulsion).
Then, the particles have energy larger than the Fermi surface by the order of $\Delta$.
When the energy is large enough, the chiral condensate is disfavored.
However, in such a situation, the exclusion effect is not so trivial, and the chiral condensate could be formed.
If any pairing between an upper mode and an antiparticle is possible, then it may imply the coexistence of chiral and Kondo condensates.
Thus, both the lower and upper modes newly induced by the appearance of the Kondo condensate are not suitable for the formation of the chiral condensate.

\subsection{At $T \neq 0$}

\begin{figure}[b!]
    \begin{minipage}[t]{1.0\columnwidth}
        \begin{center}
            \includegraphics[clip, width=1.0\columnwidth]{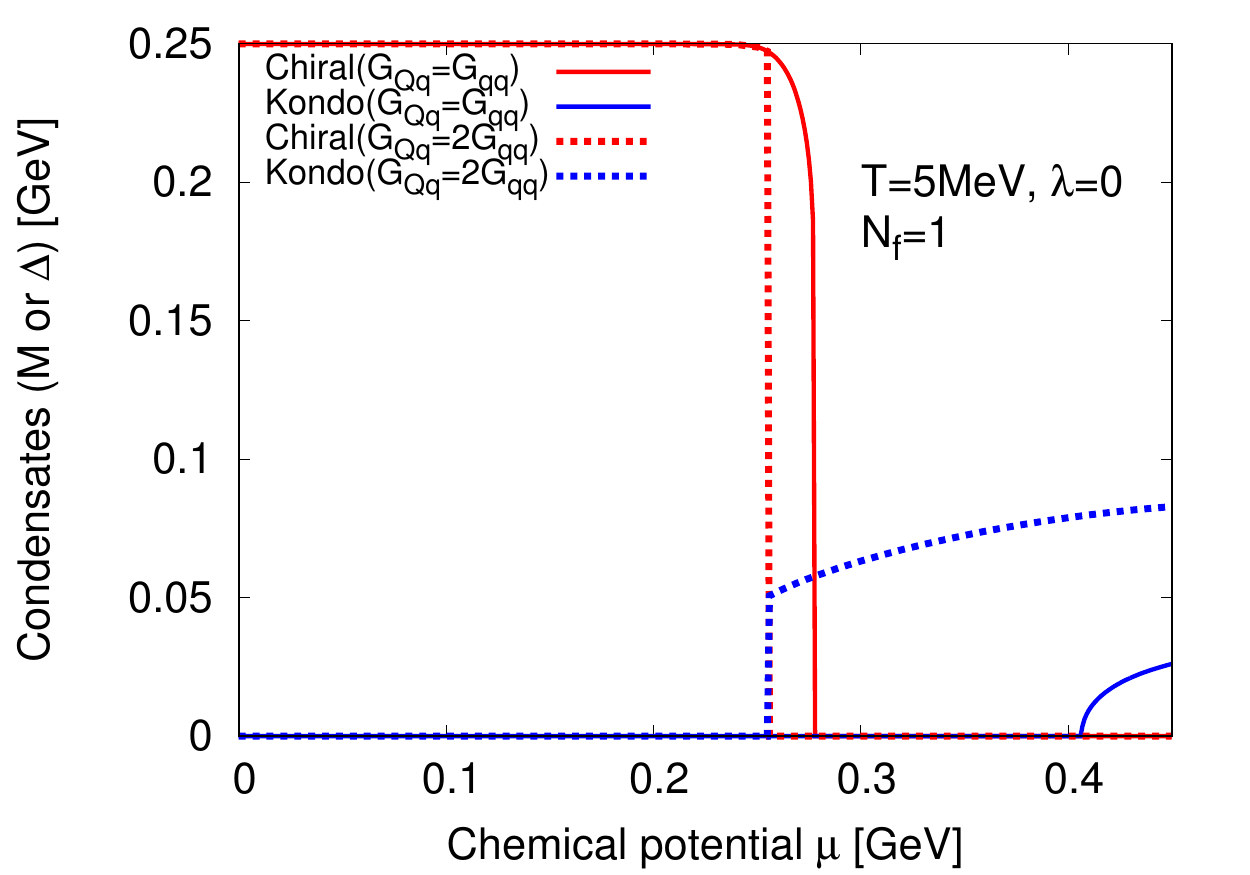}
        \end{center}
    \end{minipage}
    \begin{minipage}[t]{1.0\columnwidth}
        \begin{center}
            \includegraphics[clip, width=1.0\columnwidth]{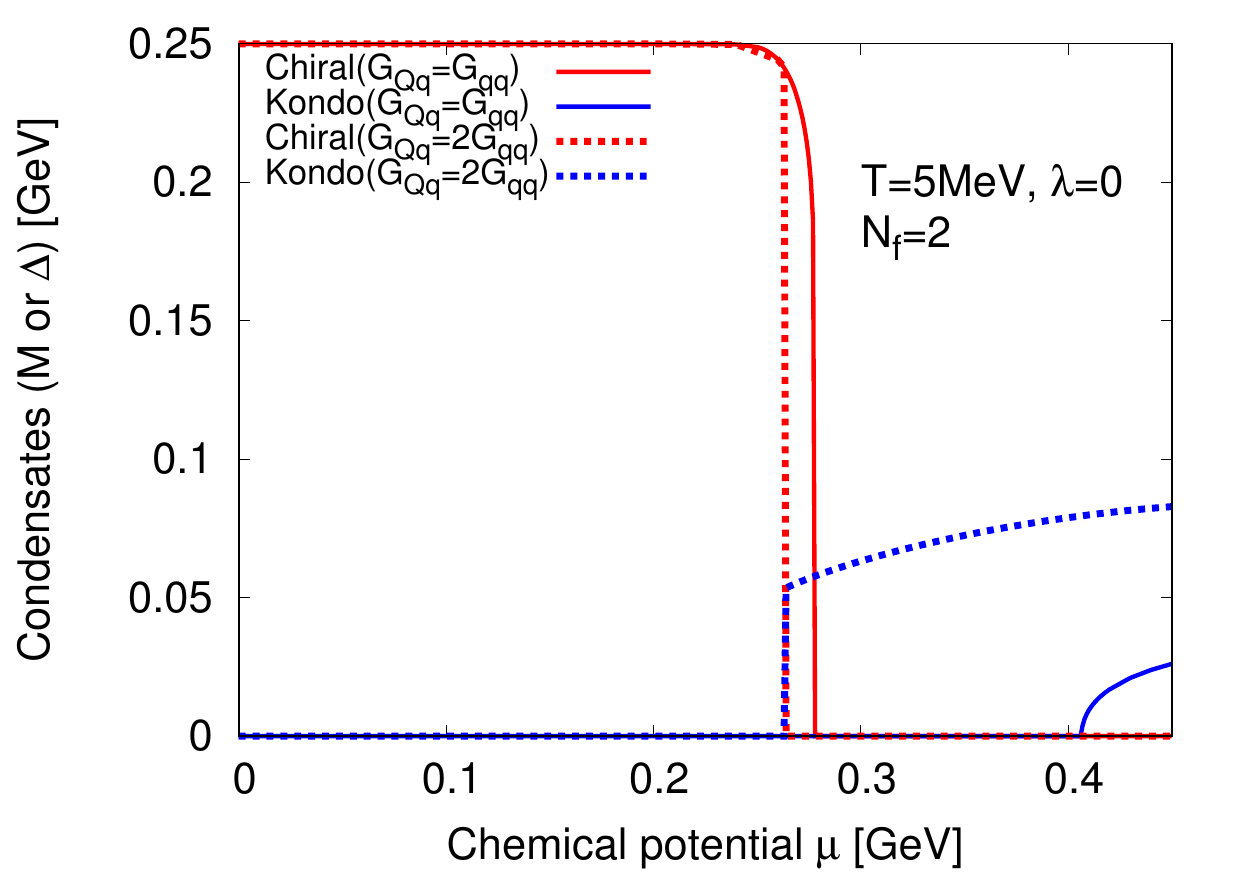}
        \end{center}
    \end{minipage}
    \caption{$\mu$ dependences of chiral and Kondo condensates at $T=5 \ \mathrm{MeV}$ and $\lambda=0$.
The light-quark flavor is set as $N_f=1$ (upper) or $N_f=2$ (lower).}
    \label{FiniteT}
\end{figure}

In this subsection, we focus on the thermal behaviors.
The results at $T=5 \ \mathrm{MeV}$ and $\lambda=0$ are shown in Fig.~\ref{FiniteT}.
Our findings are as follows:
\begin{enumerate}
\setlength{\leftskip}{1.2em}
\item The order of the transition to the Kondo phase becomes the first order due to the thermal effect \cite{Kanazawa:2016ihl}.

\item For a smaller coupling (even at $G_{Qq} = G_{qq}$), the Kondo condensate is suppressed by the thermal effect, and the onset chemical potential $\mu_c^K$ is pushed up.
The suppression of the Kondo condensate at $T=5 \ \mathrm{MeV}$ is quantitatively reasonable because the critical temperature of the Kondo condensate should be proportional to the magnitude of the gap $\Delta < 10 \ \mathrm{MeV}$.
In such a situation, $\mu_c^\chi < \mu_c^K$, and the Kondo condensate cannot contribute to the chiral condensate.

\item For a strong coupling $G_{Qq} = 2G_{qq}$, the chiral condensate can be affected by the Kondo condensate.
The exclusion of the chiral condensate for $N_f=2$ by the Kondo condensate becomes weaker than that in the $N_f=1$ case, as discussed at $T=0$.
The coexistence phase disappears due to the first-order transition of the Kondo condensate.
\end{enumerate}

\begin{figure}[t!]
    \begin{minipage}[t]{1.0\columnwidth}
        \begin{center}
            \includegraphics[clip, width=1.0\columnwidth]{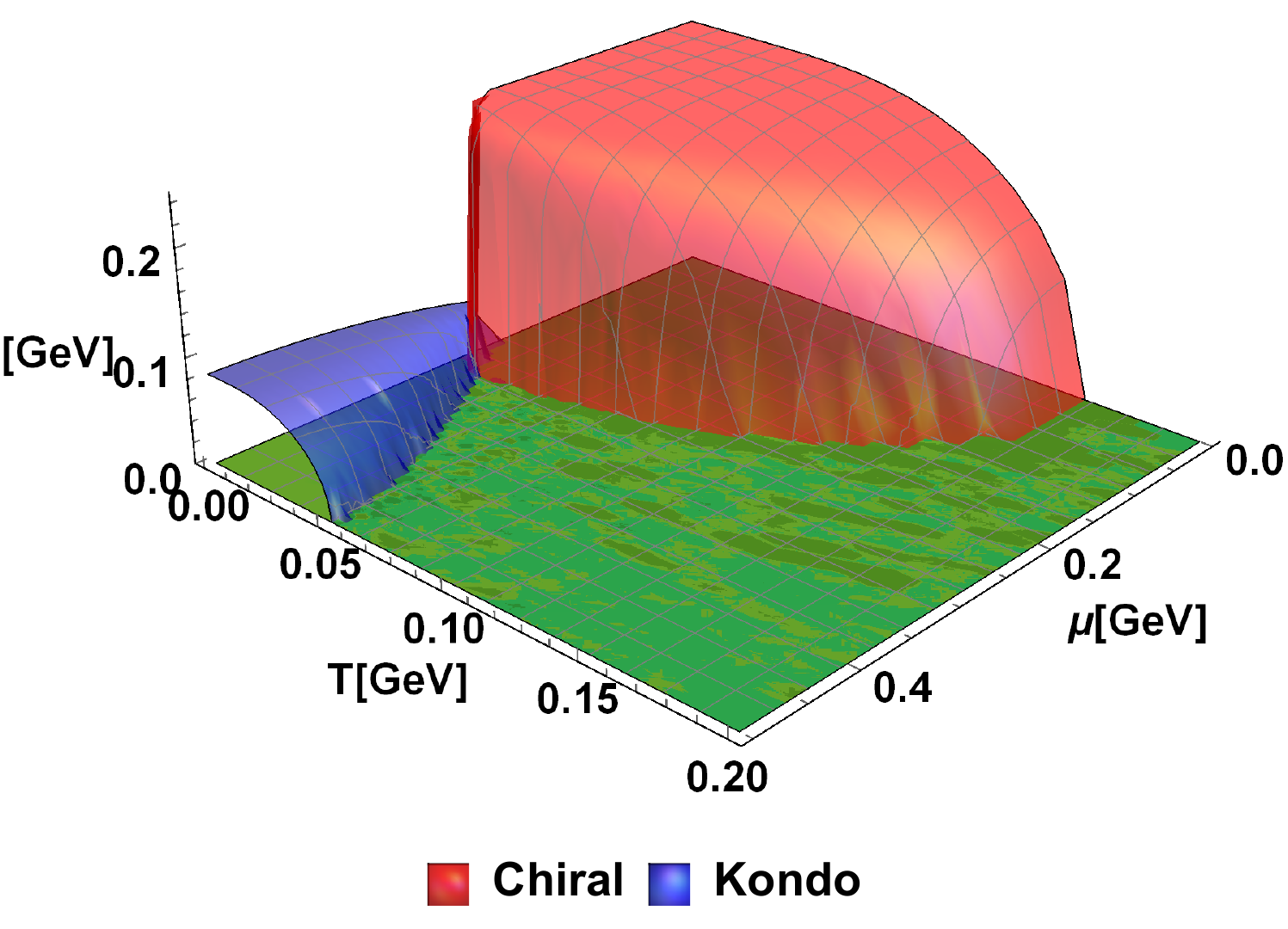}
        \end{center}
    \end{minipage}
    \caption{Phase diagram in the $T$-$\mu$ plane at $\lambda=0$ and $G_{Qq} = 2G_{qq}$ for $N_f=2$.
The red and blue regions correspond to $M>0$ and $\Delta>0$, respectively. }
    \label{T-mu-phase}
\end{figure}

Finally, we summarize the phase diagram on the $T$-$\mu$ plane at $\lambda=0$ and $G_{Qq} = 2G_{qq}$ for $N_f=2$ in Fig.~\ref{T-mu-phase}.

\subsection{At $\lambda \neq 0$}
Next we focus on the region of finite $\lambda>0$.\footnote{Also, we can check the case of $\lambda<0$, but in this case the results suffer from an artificial (or nonphysical) effect by ultraviolet cutoff $\Lambda$ \cite{Yasui:2016svc}. Although the region of $\lambda<0$ is also interesting, we focus on $\lambda>0$ in this work.}
In the (usual) light-quark matter with $\lambda = 0$, the energy necessary to generate a static heavy quark, is the heavy quark mass $m_Q$.
On the other hand, a nonzero $\lambda$ corresponds to the energy necessary to put a {\it virtual component} (or residual momentum) of a static heavy quark into the system.
Therefore, in light-quark matter with $\lambda \neq 0$, the generation of a static heavy quark participating in the Kondo condensate requires not only the heavy quark mass $m_Q$ but also an additional energy shift compensating $\lambda$.

\begin{figure}[t!]
    \begin{minipage}[t]{1.0\columnwidth}
        \begin{center}
            \includegraphics[clip, width=1.0\columnwidth]{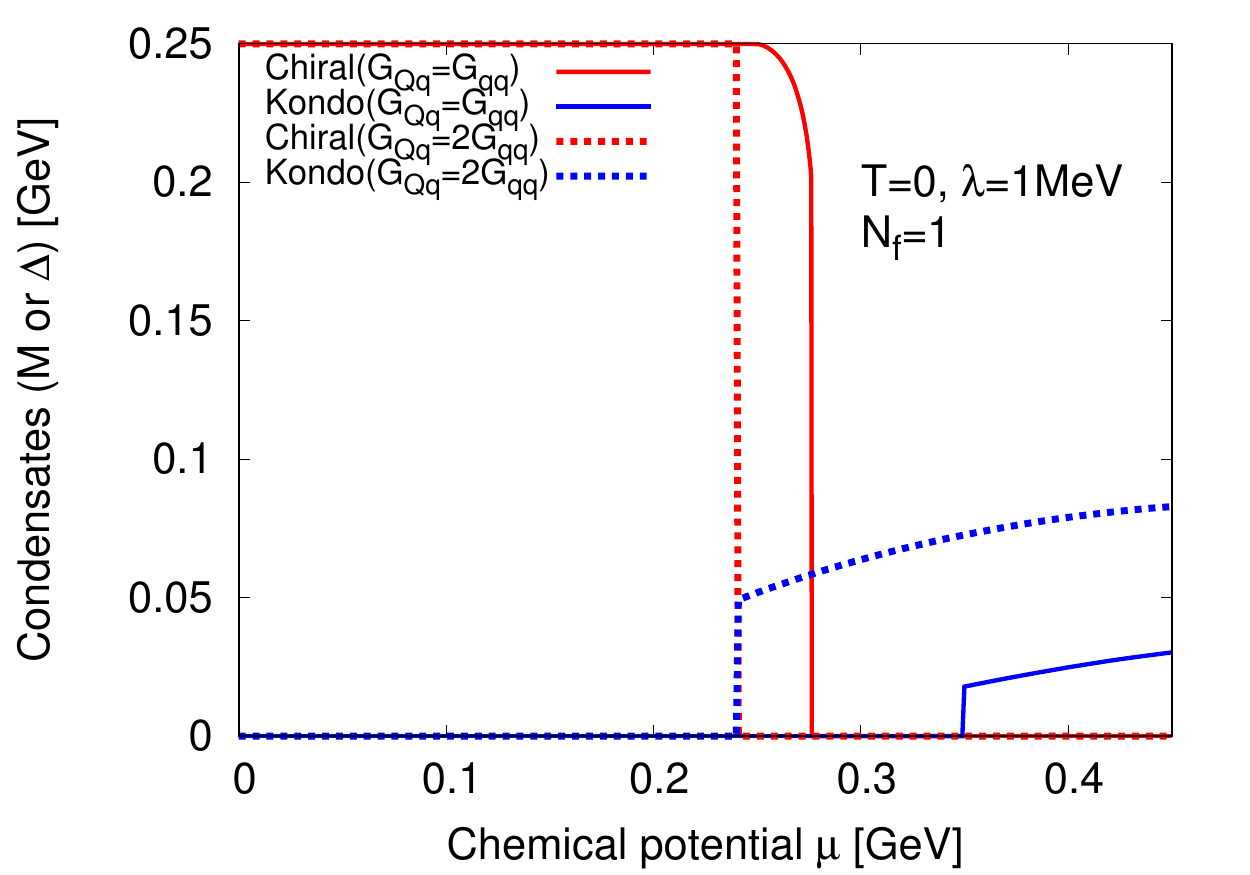}
        \end{center}
    \end{minipage}
    \begin{minipage}[t]{1.0\columnwidth}
        \begin{center}
            \includegraphics[clip, width=1.0\columnwidth]{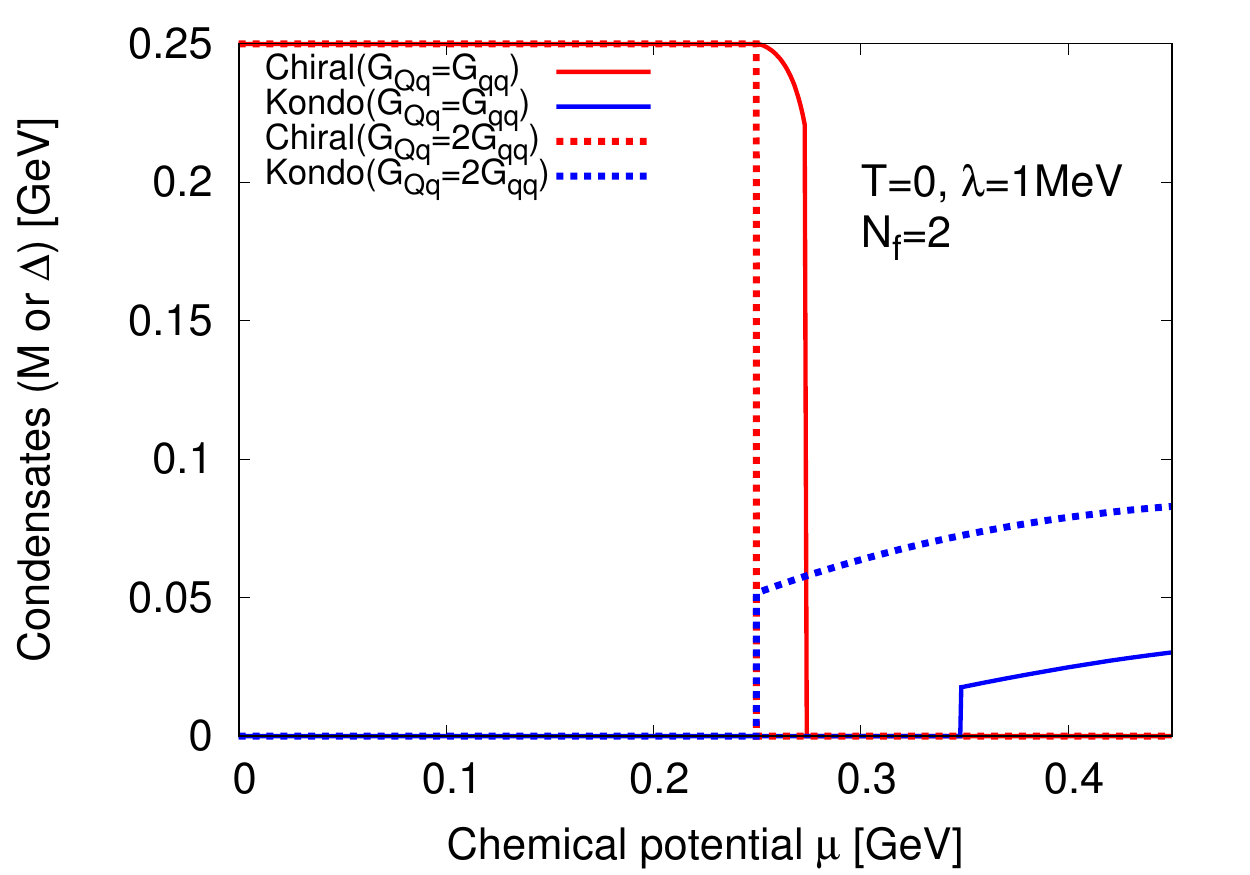}
        \end{center}
    \end{minipage}
    \caption{$\mu$ dependences of chiral and Kondo condensates at $T=0$ and $\lambda=1 \ \mathrm{MeV}$.
The light-quark flavor is set as $N_f=1$ (upper) or $N_f=2$ (lower).}
    \label{FiniteLam}
\end{figure}

The results at $\lambda=1 \ \mathrm{MeV}$ and $T=0$ are shown in Fig.~\ref{FiniteLam}.
At finite $\lambda>0$, the order of the phase transition of the Kondo condensate becomes first order \cite{Yasui:2016svc,Kanazawa:2016ihl}. 
If the Kondo condensate is small (the coupling constant $G_{Qq}$ is small enough), the critical chemical potential of the Kondo condensate becomes larger than that of the chiral phase transition, $\mu_c^\chi < \mu_c^K $.
Then, the Kondo condensate does not contribute to the chiral condensate.
The $\mu$ dependence of the onset of the Kondo condensate is slightly different than that of the finite-temperature case.
Moreover, if $G_{Qq}$ is strong enough, the Kondo condensate can affect the chiral condensate, and $\mu_c^\chi$ is pushed down by the Kondo condensate.
As a result, we see $\mu_c^\chi(N_f=1) < \mu_c^\chi(N_f=2)$, as discussed at $\lambda=0$.
The coexistence phase disappears due to the first-order transition of the Kondo condensate, which is the same situation as in the finite-temperature case.

\subsection{Current light-quark mass effect}
We investigate effects of light flavors with a finite current quark mass in the Kondo condensate.
For simplicity, let us neglect the chiral condensate and replace $M$ by the current light-quark mass $m_1$ (or $m_2$) of $\psi_1$ (or $\psi_2$).
Then, using Eqs.~(\ref{dis_Epm_Nf=1}), (\ref{dis_Etil_Nf=1}), and (\ref{dis_Epm_Nf=2})--(\ref{dis_Etil_Nf=2}), the dispersion relations of massive light flavors can be written as 
\begin{eqnarray}
E_{1\pm}(p) &\equiv&  \frac{1}{2} \Biggl( \sqrt{p^2 + m_1^2} + \lambda -\mu \Biggr. \nonumber \\
&& \left. \pm \sqrt{ \left(\sqrt{p^2 + m_1^2}-\lambda-\mu \right)^2 + 8 |\Delta|^2 }\right), \label{dis_m1pm} \\
E_2(p) &\equiv& \sqrt{p^2 + m_2^2} - \mu, \label{dis_m2}\\
\tilde{E}_1 (p) &\equiv& - \sqrt{p^2 + m_1^2} - \mu, \label{dis_m1til} \\
\tilde{E}_2 (p) &\equiv& - \sqrt{p^2 + m_2^2} - \mu. \label{dis_m2til}
\end{eqnarray}
These dispersion relations are put into the thermodynamic potential (\ref{omega}): we use Eqs.~(\ref{dis_m1pm}) and (\ref{dis_m1til}) for $N_f=1$, and Eqs.~(\ref{dis_m1pm})--(\ref{dis_m2til}) for $N_f=2$.

\begin{figure}[t!]
    \begin{minipage}[t]{1.0\columnwidth}
        \begin{center}
            \includegraphics[clip, width=1.0\columnwidth]{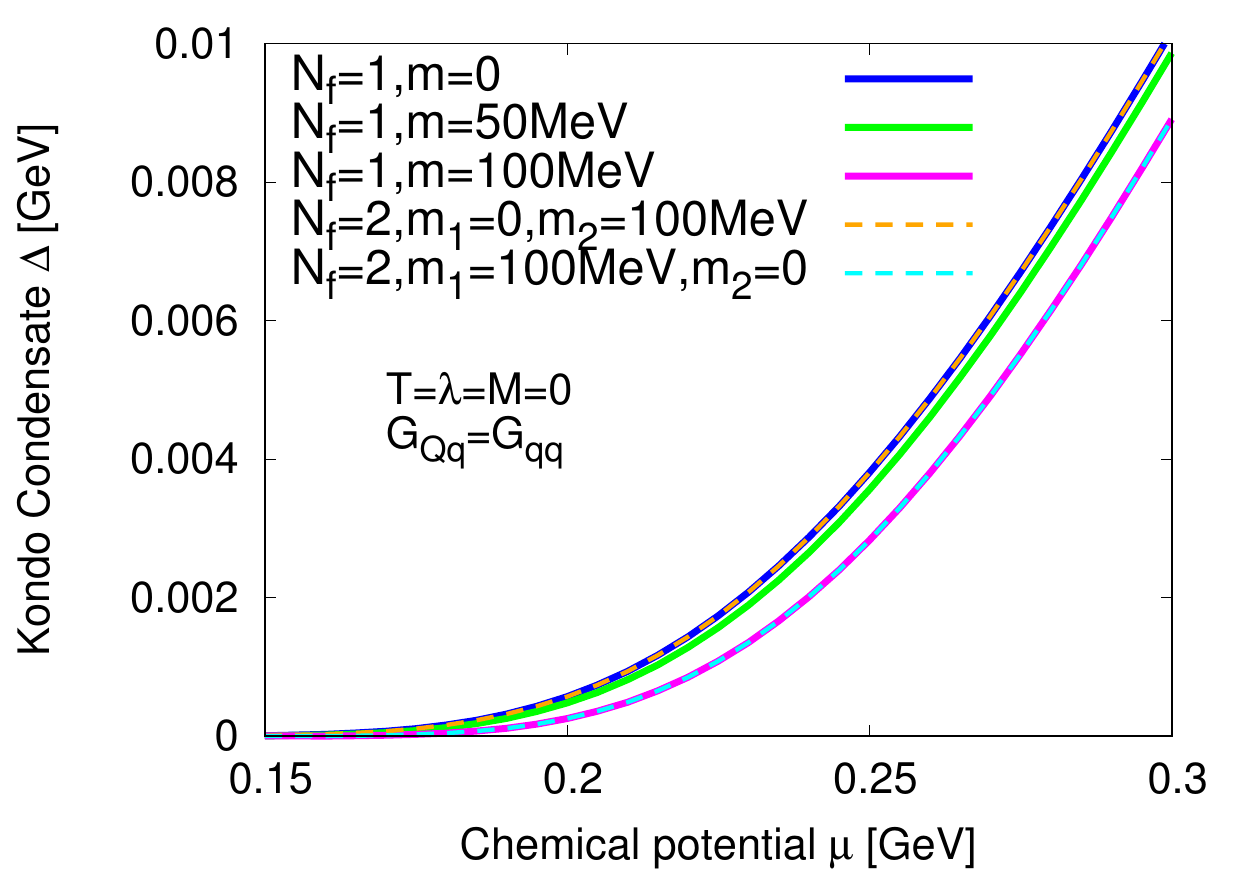}
        \end{center}
    \end{minipage}
    \caption{Current light-quark mass dependences of the Kondo condensate at $M=T=\lambda=0$ and $G_{Qq} = G_{qq}$.}
    \label{mass_dep}
\end{figure}

The numerical results are shown in Fig.~\ref{mass_dep}.
Our findings from this figure are as follows:
\begin{enumerate}
\setlength{\leftskip}{1.2em}
\item The inclusion of a current light-quark mass suppresses the magnitude of the Kondo condensate.
Such an effect is the same as that of the chiral condensate.

\item The effect of the current light-quark mass is quite small for up and down quarks with $m_{u,d} <10 \ \mathrm{MeV}$, and even strange quarks with $m_s \sim 100 \ \mathrm{MeV}$ can form the Kondo condensate.
Such a conclusion might be important in the sense that, in the two-flavor color-superconducting phase with heavy-quark impurities, strange quarks which do not participate in the (light) quark-quark pairing can induce the Kondo condensate due to the coupling to heavy quarks. 

\item For $N_f=2$, we checked some patterns of $(m_1,m_2)$.
We found that the result for $(m_1=0,m_2=100 \ \mathrm{MeV})$ agrees with that of $N_f=1$ at $m_1=0$.
On the other hand, the result for $(m_1=100 \ \mathrm{MeV},m_2=0)$ is the same as that of $N_f=1$ at $m_1=100 \ \mathrm{MeV}$.
This is because we assumed that only the first light flavor is coupled to the heavy flavor.
The second flavor does not couple to the heavy flavor [see Eq.~(\ref{detNf=2})], so its current mass never contributes to the Kondo condensate.

\item Including a finite light-quark mass does not change the order of the phase transition with increasing $\mu$.
This situation is the same as that of the chiral condensate.
\end{enumerate}

\section{Summary and outlook} \label{Sec_4}
In this paper we investigated how the heavy-light quark condensate induced by the QCD Kondo effect affects the chiral condensate.
As a result, we found a decrease of the critical chemical potential for the chiral phase transition and the coexistence phase of Kondo and chiral condensates.

We emphasize that our analyses indicate that {\it the nonperturbative region of the QCD Kondo effect can modify the properties of the chiral condensate} near the critical density.
These effects can affect the transport and thermodynamic observables in high-density matter with heavy quarks/hadrons, such as that created by low-energy heavy-ion collisions at FAIR, NICA, and J-PARC.
In particular, since $D$, $D_s$, and $B$ mesons are promising probes of chiral symmetry restoration at high density \cite{Hilger:2008jg,Blaschke:2011yv,Hilger:2011cq,Sasaki:2014asa,Buchheim:2014rpa,Suenaga:2014sga,Suzuki:2015est,Park:2016xrw,Harada:2016uca,Suenaga:2017deu,Hosaka:2016ypm}, the competition with the Kondo condensate would play an important role in their observables.

We comment on the coupling constants of our model, $G_{Qq}$ and $G_{qq}$.
In this study, we checked the three cases of $G_{Qq}>G_{qq}$, $G_{Qq}=G_{qq}$, and $G_{Qq}<G_{qq}$ with $G_{qq}$ fixed in vacuum.
Among them, we showed that the larger heavy-light coupling can most clearly induce visible interplay effects.
To go beyond the current approach, one may consider $\mu$-dependent coupling constants,
but this would lead to serious artifacts of the model with the four-point interactions.
A more careful treatment of the interaction vertices at finite density will be needed for more quantitative studies,
where the gluon exchange should be considered as explicit degrees of freedom in the interaction between two quarks.

Also, we comment that the color-superconducting phase is important for investigating the Kondo condensate.
In Ref.~\cite{Kanazawa:2016ihl}, a competition between the Kondo and color-superconducting phases was investigated, and it was found that the Kondo condensate will be suppressed by a paring of light quarks.
However, it should be noted that the magnitude of the color-superconducting gap depends on the diquark coupling constant, which is not so reliable in effective models, and further detailed investigations are required in the future.
For example, in Ref.~\cite{Buballa:2003qv} the color-superconducting gap was known to be $\Delta_{\mathrm{CS}}=140 \ \mathrm{MeV}$ at $\mu=500 \ \mathrm{MeV}$, which is comparable with the magnitude of the Kondo condensate in this work.
In particular, near the critical chemical potential, an investigation of the competition (or coexistence) of the Kondo, chiral, and diquark condensates would also be interesting.

\section*{Acknowledgments}
This work is partially supported by the Grant-in-Aid for Scientific Research (Grants No.~25247036, No.~15K17641, No.~16K05366, No.~17K05435, and No.~17K14277) from the Japan Society for the Promotion of Science.
K. S. is supported by MEXT as ``Priority Issue on Post-K computer" (Elucidation of the Fundamental Laws and Evolution of the Universe) and JICFuS.

\bibliography{reference}

\end{document}